\documentclass{article} 

\usepackage{graphicx} 
\usepackage{caption}
\usepackage{subcaption}
\usepackage{amsmath}
\usepackage{amssymb}
\usepackage{capt-of}
\usepackage{float}
\usepackage{wrapfig}

\usepackage{mathtools}
\DeclarePairedDelimiter\floor{\lfloor}{\rfloor}

\usepackage{authblk}

\graphicspath{{./images/}} 

\usepackage[numbers]{natbib}
\def\name#1{\vskip5pt\par{{\sc #1}}}
\def\address#1{\par{\textit{#1}}}
\def\email#1{\par{\rm #1}}

\begin{document}


\title{Planar growth generates scale-free networks}


\author{%
\name{Garvin Haslett$^{1}$ \textsuperscript{\textdagger}, Seth Bullock$^2$ and Markus Brede$^1$}
\address{$^1$ Institute for Complex Systems Simulation, School of Electronics and Computer Science, University of Southampton SO17 1BJ, UK}
\address{$^2$ Department of Computer Science, University of Bristol BS8 1UB, UK}
\email{\textsuperscript{\textdagger} Corresponding author email: g.a.haslett@soton.ac.uk}}
\maketitle

\begin{abstract}
{In this paper we introduce a model of spatial network growth in which nodes are placed at randomly selected locations on a unit square in \(\mathbb{R}^2\), forming new connections to old nodes subject to the constraint that edges do not cross. The resulting network has a power law degree distribution, high clustering and the small world property. We argue that these characteristics are a consequence of the two defining features of the network formation procedure; growth and planarity conservation. We demonstrate that the model can be understood as a variant of random Apollonian growth and further propose a one parameter family of models with the Random Apollonian Network and the Deterministic Apollonian Network as extreme cases and our model as a midpoint between them. We then relax the planarity constraint by allowing edge crossings with some probability and find a smooth crossover from power law to exponential degree distributions when this probability is increased.}

\end{abstract}


\newpage

\section{Introduction}

The field of spatial networks is emerging as an important topic within network science \cite{barthelemy2011spatial}. The distinguishing feature of this work is that network nodes are assigned a position in Euclidean space, typically \(\mathbb{R}^2\), with the distance between them described by the Euclidean metric. A major goal in this area is to investigate the way in which constraining connectivity in a manner related to node proximity influences network organisation. Application domains include city science \citep{cardillo2006structural, xie2007measuring, jiang2007topological, barthelemy2008modeling, masucci2009random, chan2011urban, courtat2011mathematics, strano2012elementary, levinson2012network, rui2013exploring, gudmundsson2013entropy}, electronic circuits \citep{i2001topology, bassett2010efficient, miralles2010planar, tan2014global}, wireless networks \citep{huson1995broadcast, lotker2010structure}, leaf venation \citep{corson2010fluctuations, katifori2010damage}, navigability \citep{kleinberg2000navigation, PhysRevLett.108.128701, huang2014navigation} and transportation \citep{gastner2006shape, louf2013emergence}. Spatial networks vary in the extent to which they respect \emph{planarity}: the property of a spatial network having edges that do not cross. While, for example, sexual contact networks may be embedded in space, they need not respect planarity. By contrast, the layout of a microchip must be planar since conductor lines may not cross without creating a junction. Transport networks tend to be nearly planar (a relatively small number of bridges and tunnels allow edges to cross without creating a junction vertex). Planarity is also a consideration in the construction of infrastructure such as wireless networks \citep{cairns2013revisiting}. Despite the relevance of planarity considerations across a wide range of nework domains, the role of planarity in network formation is an under-represented issue in the spatial networks literature \citep{newman2010networks}.

Following Barab{\'a}si \& Albert's demonstration that preferential attachment results in a scale-free network \cite{barabasi1999emergence}, the conditions under which the power-law degree distribution obtains within spatial networks has been a significant area of investigation; for a review see reference \cite{hayashi2006review}. Three principal classes of mechanism have been identified in this regard \cite{hayashi2006review}; (i) link length penalisation, (ii) embedding a scale-free network within a lattice and (iii) space filling. Two models make up class (i); the modulated BA model and the geographical threshold graph. The first of these two is an extension of the BA model where the probability of a new connection is inversely proportional to the euclidean distance between the nodes under consideration and the degree of the existing node. As such, the modulated BA implicitly assumes a power law degree distribution. In the geographical threshold model, nodes are connected when the product of their respective weights and a function of the distance between them exceeds a pre-defined constant. In this case a scale-free network results only when either the distribution of the weights or the distance function follows a power law. Furthermore, we note that the canonical model of link length penalisation, the random geometric graph, only produces a scale-free degree distribution when the nodes are themselves inhomogeneously distributed on the plane \cite{herrmann2003connectivity, barnett2007spatially, bullock2010spatial}. Models in class (ii) assign an intrinsic degree \(k\) to all nodes in a lattice. Members of the lattice are then selected at random and connected to their \(k\) nearest neighbours subject to a distance constraint \cite{ben2003geographical, rozenfeld2002scale}. The degree distribution in this case is precisely that which is assigned to the model during its construction. Thus, models in classes (i) and (ii) only result in a scale-free distribution when a power law is assumed as some aspect of their inputs. 

Class (iii) recursively partitions the space by adding new nodes to the plane and then connects them to the existing graph. Two of these models, the Apollonian network (hereafter DAN, the Deterministic Apollonian Network) \cite{doye2004self, andrade2005apollonian} and its stochastic variant the Random Apollonian Network (hereafter RAN) \cite{zhou2005maximal}, are of primary interest to this study. Both models choose faces of an existing triangulation of the plane and split them into three, resulting in a new triangulation. Analytical treatment of their respective degree distributions reveal them to be power laws of the form \(P(k) \sim k^{-\alpha}\). For the DAN, this exponent is \(\alpha_{DAN} = 1 + \text{ln }3 / \text{ln }2 \approx 2.585\) and for the RAN it is \(\alpha_{RAN} = 3.0\). Further work in this vein has investigated the average path length \cite{zhang2008analytical}, degree spectrum \cite{andrade2005spectral} and dynamical properties  of the DAN \cite{pellegrini2007activity, buesser2012evolution} while a similar body of research exists for the RAN \cite{frieze2012certain, zhang2014number}. A unifying framework for Apollonian networks, the Evolutionary Apollonian Network, is a triangulation model which can be induced to produce either the DAN or the RAN by variation of a single parameter \cite{zhang2006evolving, kolossvary2013degrees}. However, none of these Apollonian growth models attribute an explicit point in space to their nodes; in each case it is network topology that determines the outcome of the process. Furthermore, when interpreted spatially, the nodes of these models are not distributed uniformly in space.

As a final example in class (iii), we highlight the model of Mukherjee \& Manna \cite{mukherjee2006weighted}. Here, new nodes are connected to a random end of the nearest edge. The model is notable in that it is the only existing spatial growth process we have identified that results in a scale-free distribution when (a) nodes are distributed uniformly at random on the plane and (b) there are no other inputs to the model that have a power law form.

In this paper we present two related mechanisms; the first, planar growth (PG), seeks to directly address the impact of a planarity constraint on a network growth process. Briefly, PG incrementally builds a network by placing new nodes at random locations in space and connecting them to other nodes such that planarity is maintained. We introduce it in section \ref{sec:models} alongside two reference cases; one of which considers a network that grows in time but does not enforce planarity, while the other considers a network built over a static set of nodes through the addition of planarity-preserving edges. In contrast to the reference cases, PG results in a power law degree distribution and we present evidence to support this claim in section \ref{sec:degree}. Further investigation of PG is presented in sections \ref{sec:analysis} and \ref{sec:planar:relax_planarity}, where we examine other key network measures and demonstrate the consequences of relaxing planarity, respectively.  

The second mechanism is named Apollonian Planar Growth (APG) and is introduced in section \ref{sec:apg} as a reformulation of PG as an Apollonian growth process. Consideration of APG as an object of study in its own right leads to further contributions. Firstly, the APG is inherently spatial; in contrast with the topological character of its precursors, the DAN and the RAN. Secondly, PG can be viewed as a generalisation of Apollonian growth processes to cases where \(m\), the number of connections made when a node is added to the network, is less than 3. In section \ref{sec:apg:comparison}, we further develop APG as a single parameter model, the variation of which tunes the exponent of the network's degree distribution. The DAN and the RAN can then be seen as special cases of APG, with PG intermediate between them. Finally, we conclude this paper in section \ref{sec:summary} where we summarise our results.

\section{The models and their degree distributions}

We begin with a description of the models and an analysis of the degree distributions that they produce.

\label{sec:models}
\subsection{Planar Growth, no planarity and no growth}

Planar Growth creates spatially embedded networks with \(N + 10\) nodes and average degree \(2m\) on a unit Euclidean square that has rigid boundary conditions. We wish to begin the process with a planar network that has nodes distributed uniformly on the plane. To do so ten nodes are placed uniformly at random upon the unit square with \(m \times 10\) planar edges between them. As they are added, each node after the first is connected to an existing node; the edge being chosen so as not to violate planarity. Once all ten nodes have been placed, unconnected pairs are then chosen at random and an edge is chosen between them; again subject to the caveat that planarity is always maintained. We continue choosing node pairs until \(m \times 10\) edges are added or until all possible node pairs have been tried. The resulting network is accepted irrespective of its final number of edges.

Tests of the procedure over 10,000 realisations show that for \(m = 2\) the average degree of the initial network was \(k_{ave} = 3.99\), while for \(m = 3\) it was \(k_{ave} = 4.2\). Despite the results for the \(m = 3\) case we retain this method of initialisation since the number of nodes and edges is statistically insignificant in comparison to the finished network and the method reliably produces initial networks with the desired properties.

The algorithm now enters the growing phase where the following steps are repeated \(N\) times: \\

\noindent (1) Place a new node, \(i\), uniformly at random within the square.
        
\noindent (2) Repeat \(m\) times:

  (2a) Pick a node \(j\) where \(j \neq i\).
        
  (2b) If \(i\!j\) does not cross an existing edge then add \(i\!j\) otherwise go to 2a. \\
        
\noindent If step 2 cannot be completed because \(m\) valid nodes do not exist then remove node \(i\) and any associated edges and repeat step 1. \\

As reference cases for PG we consider two degenerate variants of the mechanism; one with no planarity constraint, \(\text{PG-noplanarity}\), and one with no growth, \(\text{PG-nogrowth}\). \(\text{PG-noplanarity}\) is very similar to PG except that edge connections are always allowed. This scenario is equivalent to the uniform attachment model originally introduced by Barab{\'a}si \& Albert \cite{barabasi1999mean} where it was shown to result in networks with an exponential degree distribution. In \(\text{PG-nogrowth}\) we create a static population of \(N\) nodes placed uniformly on the unit square. Pairs of nodes are picked at random and an edge is drawn between them, provided this new edge does not cross an existing one. We continue until \(N \times m\) edges have been added.

\subsection{Analysis of the degree distribution}
\label{sec:degree}

\begin{figure}
  \begin{center}
    \includegraphics[width=0.95\textwidth]{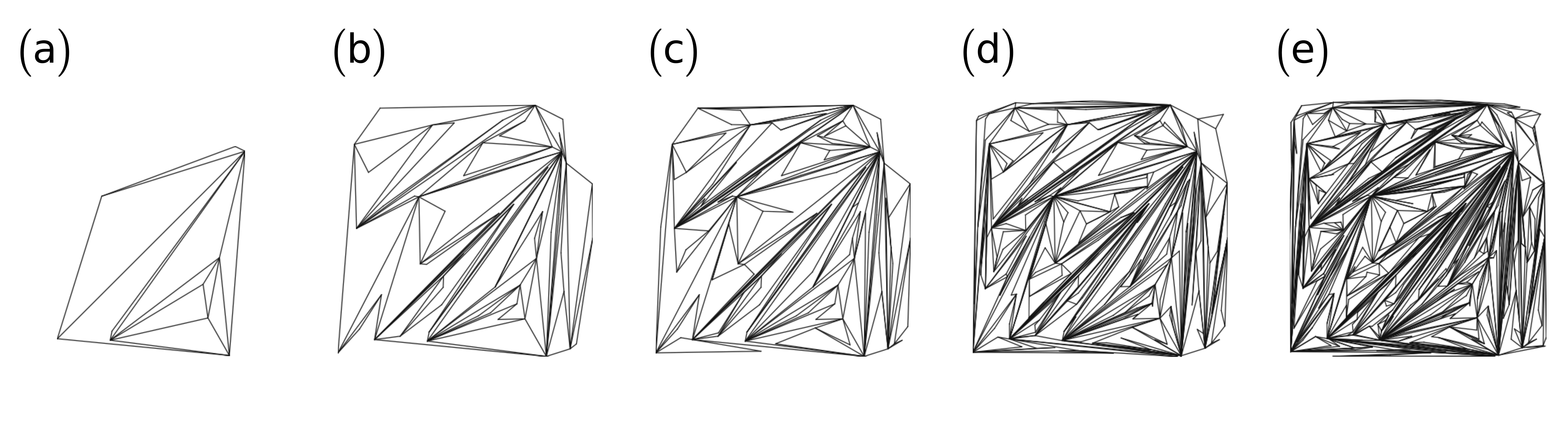}
  \end{center}
  \caption{A PG network with \(m=2\) at various stages of its growth. (a) \(N = 0\) (b) \(N = 50\) (c) \(N = 100\) (d) \(N = 250\) (e) \(N = 500\).}
  \label{fig:growing}
\end{figure}
\begin{figure}
  \begin{center}
    \includegraphics[width=0.9\textwidth]{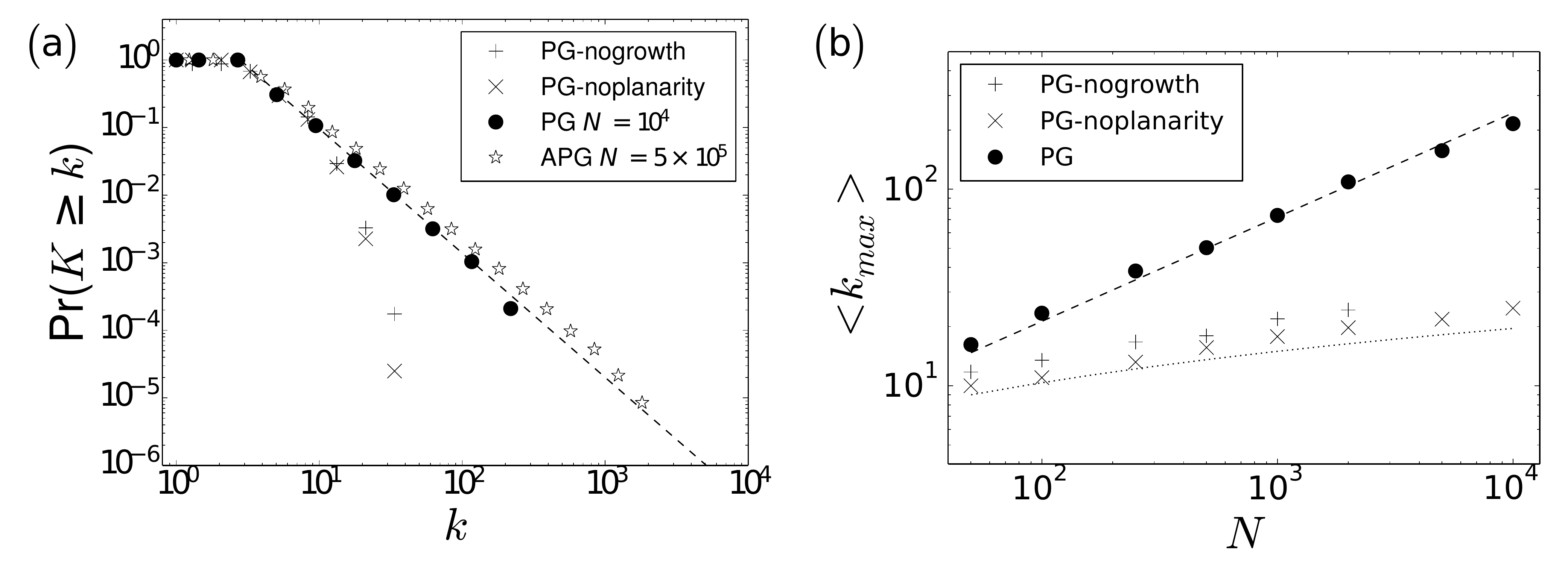}
  \end{center}

  \caption{(a) Cumulative degree distributions for PG networks of order \(N = 10^4\), APG networks of order \(5 \times 10^5\), \(\text{PG-noplanarity}\) networks of order \(N = 10^4\) and \(\text{PG-nogrowth}\) networks of order \(N = 2 \times 10^3\). All results averaged over 20 experiments with \(m = 2\). The dashed line is the best fit for the APG experiment, a power law with exponent, \(\alpha_{\text{APG}} = 2.77 \pm 0.01\). As with all exponents in this paper, \(\alpha_{\text{APG}}\) has been estimated using the method of Maximum Likelihood Estimators outlined in Clauset et al.\ \cite{clauset2009power}. (b) Average maximum degree observed for the PG, \(\text{PG-noplanarity}\) and \(\text{PG-nogrowth}\) networks. The dashed line is the expected value of the maximum degree for a power law with exponent \(\alpha_{m=2} = 2.83 \pm 0.01\), the estimated value of the exponent in the \(n = 10^4, m = 2\) case. The dotted line is a plot of the expected maximum degree for a network with an exponential distribution.}
  \label{fig:m2_mechanisms_k_hist}
\end{figure}
Figure \ref{fig:growing} is a series of visualisations of a PG network from its initialisation until it reaches 500 nodes. Qualitatively it seems that some nodes acquire a disproportionately high amount of connections hinting that the network has a skewed degree distribution. We proceed, in figure \ref{fig:m2_mechanisms_k_hist}a, with a plot of the degree distribution for a planar growth experiment of order \(N = 10^4\), along with a \(\text{PG-noplanarity}\) experiment of order \(N = 10^4\) and a \(\text{PG-nogrowth}\) experiment of order \(N = 2 \times 10^3\). A smaller value of \(N\) is reported for \(\text{PG-nogrowth}\) due to computational limits. Nonetheless the results show the degree distributions of both reference cases to be exponential while the PG experiment approximates a power law distribution.

To investigate finite size effects we plot, in figure \ref{fig:m2_mechanisms_k_hist}b, how the maximum degree observed during these experiments varies with the size of the network. Following Newman \cite{newman2003structure}, we also plot the analytically derived relationship between \(\langle k_{max} \rangle\), the mean maximum degree for networks with a power law degree distribution, and \(N\); \(\langle k_{max} \rangle \sim N^{1/(\alpha - 1)}\). We find it to be in good agreement with the observations which provides strong support for the hypothesis of a power law distribution.

The expected value of the \(i^{\mathrm{th}}\) member of a sequential ordering of the random variables of an exponential distribution with parameter \(\lambda\) is \(E[X_i] = H_i/\lambda\), where \(H_i\) is the \(i^{\mathrm{th}}\) harmonic number. Barab{\'a}si \& Albert found the degree distribution for the uniform attachment model to be \(P(k) = e\lambda \text{ exp}(-\lambda k)\) with \(\lambda = 1/m\). We therefore approximate the average maximum degree for a \(\text{PG-noplanarity}\) network of order \(N\) with \(\langle k_{max} \rangle \sim mH_N \). The plot of this curve also matches well with our empirical data supporting the claim that the degree distributions generated for both types of reference cases are exponential.

Considered as a whole, the evidence in this section suggests that the network produced by the planar growth process is scale-free. The necessary ingredients in order to produce this outcome are growth and the planarity conservation. When either of these aspects are removed we observe an exponential degree distribution. However, results discussed in this section are unsatisfactory in that the distribution has only been shown to hold over one order of magnitude. We will attend to this in the next section.

\subsection{Apollonian Planar Growth}
\label{sec:apg}

Zhou's original RAN algorithm \cite{zhou2005maximal} starts with an equilateral triangle on the plane. Network construction proceeds by repeatedly choosing a face of the triangulation at random, placing a new node within it and connecting that node to the vertices of the face. Note that the probability of a node receiving a new edge is proportional to the number of triangles of which it is a vertex. This count of triangles is, in turn, equal to the degree. As such, the RAN is a form of linear preferential attachment; furthermore, its degree distribution can be analytically demonstrated to be a power law with exponent \(\alpha_{\text{RAN}} = 3.0\) when the degrees of the three vertices of the external triangle are ignored.

Apollonian Planar Growth (APG) refines this algorithm by giving the nodes an explicit position on the face of the triangle. Which face is chosen to receive a new node is still random but now this probability is in proportion to the area of the face, i.e., face \(i\) is chosen with probability \(\pi_i\) defined by the following formula:

\begin{align}
\pi_i(t) = \frac{a_i}{\sum\limits_{j \in F_t} a_j},
\label{eqn:apg}
\end{align}

\noindent where \(a_i\) is the area of face \(i\) and \(F_t\) is the set of faces present in the simulation at step \(t\).

The new node is then placed uniformly at random within triangle \(i\) and connected to its vertices. Clearly, this algorithm is equivalent to planar growth on a triangle with \(m = 3\). It has the advantage that the triangulation can be represented as a ternary tree \cite{albenque2008some}, thereby allowing for more efficient implementation of the model. Thus, in figure \ref{fig:m2_mechanisms_k_hist}a we present a plot of the degree distribution of an APG network of \(5 \times 10^5\) nodes which shows the fit of the power law extending over two orders of magnitude on both axes with an estimated exponent of \(\alpha_{\text{APG}} = 2.77 \pm 0.01\).

\subsection{Robustness to variation of \(m\)}

\label{sec:robust}

We now vary \(m\), the number of connections introduced with each new node, to determine if our observations are peculiar to the \(m = 2, 3\) cases. Three is an upper bound on \(m\), which can be established by consideration of Euler's formula for a planar graph, see discussion in reference \cite{barthelemy2011spatial} for details. We therefore vary \(m\) between one and three. Non-integer values of \(m\) are attained by always attaching \(\floor*{m}\) edges to a new node and then attaching a further node with probability \(m - \floor*{m}\). The network size in these experiments was fixed at \(n = 10^4\) and were observed to exhibit power laws. In table \ref{tbl:m_alpha} we report the estimated exponents for these networks which decrease from \(\alpha_{m = 1} = 3.15 \pm 0.03\) to \(\alpha_{m = 3} = 2.69 \pm 0.01\). From this point of view PG can be thought of as a generalisation of APG, which strictly has \(m = 3\), to any average degree less than three.

\subsection{Statistical test of the power law hypothesis}
\label{sec:stat_test}


In this section we have estimated several different exponents of assumed power law distributions using the method of Maximum Likelihood Estimation introduced by Clauset et al.\ \cite{clauset2009power}. MLE can be used as a principled method to estimate the exponent but does not establish if a power law is an appropriate model to describe the data under consideration. To do so Clauset et al.\ describe two further steps; firstly, goodness of fit is quantified by a \(p\)-value calculated by bootstrapping from the estimated model and comparing using the Kolmogorov-Smirnov statistic. Secondly, the power law is compared with other candidate distributions via log likelihood ratios.

We acknowledge that noise in empirical data can cause it to fail the bootstrapping test, thereby rendering the first step of Clauset et al.'s method inconclusive. We therefore follow the approach recommended by Alstott et al.\ \cite{alstott2014powerlaw} and use the second step as a means to identify the most appropriate distribution. In table \ref{tbl:m_loglike}, we report the log likelihood ratios, \(\mathcal{R}\), and associated \(p\)-values for two experiments, the PG network with \(N = 10^4, m= 2\) and the APG network of order \(N=5 \times 10^5\). The alternative distributions considered were the exponential:

\begin{align}
P(k) = Ce^{-\lambda k}
\end{align}

\noindent
the stretched exponential:

\begin{align}
P(k) = Ck^{\beta - 1}e^{-\lambda k^{\beta}}
\end{align}

\noindent
powerlaw with cutoff: 

\begin{align}
P(k) = Ck^{-\alpha}e^{-\lambda k}
\end{align}

\noindent
and lognormal:
\begin{align}
P(k) = Ck^{-1}\text{exp}\bigg[{-\frac{(\text{ln} \:  k - \mu)^2}{2\sigma^2}\bigg]}
\end{align}

\noindent
where $\alpha$, $\beta$, $\lambda$, $\sigma$ \& $\mu$ are the parameters to be estimated for the given distribution, $C$ is a constant that is dependent on these parameters and $k$ is the degree.

The power law model is favoured with high significance over the  exponential and stretched exponential models for both PG and APG networks. The lognormal model is not found to be a significantly better fit than the power law model for both network models (indicated by the high \(p\)-values). The power law with cutoff model is found to be a significantly better fit than the power law model (and also a significantly better fit than the log normal model: with \(\mathcal{R} = 3.0, p = 0.02\) for the APG network and \(\mathcal{R} = 5.1, p \sim \mathcal{O}(10^{-8})\) for the PG network). This might be expected given that it employs more parameters. Moreover, the estimated parameters for the functional form of the power law with cutoff suggest that the cutoff is not substantive. We observe that \(\alpha = 2.76, \lambda = 7.80 \times 10^{-5}\) for the APG network and \(\alpha = 2.77, \lambda = 0.0023\) for the PG network. We also note that the maximum degree observed for the APG network across all 20 experiments was \(k_{max} = 5726\) while for the PG network it was \(k_{max} = 271\). Both of these values are less than \(\lambda^{-1}\) indicating that while the cutoff may fit the data more appropriately the magnitude of the cutoff does not significantly impact the power law.

%

\begin{table}
    
    \caption{Estimated exponents of networks of order \(n = 10^4\) with varying \(m\).} 
    \begin{center}
    \begin{tabular}{| p{0.25cm}| p{0.5cm} | p{0.5cm} | p{0.5cm} | p{0.5cm} | p{0.5cm} |} \hline
    \textbf{\(m\)}      & 1 & 1.5 & 2 & 2.5 & 3 \\ \hline
    \textbf{\(\alpha\)} & 3.15 & 2.97 & 2.83 & 2.78 & 2.69 \\ \hline 
    \textbf{\(\sigma\)} & 0.03 & 0.02 & 0.01 & 0.02  & 0.01 \\ \hline

    \end{tabular}
    \end{center}

    \label{tbl:m_alpha}
    
    \bigskip
    
\small
Each exponent, \(\alpha\), is calculated from a batch of twenty experiments that grow a network of order \(n = 10^4\) using \(m\) as specified in the first row. Following Clauset et al.\ \cite{clauset2009power} we use the standard error, \(\sigma\), as our estimate of the uncertainty in power laws presented in this paper. For all other estimates of uncertainty we use the standard deviation.
\end{table}

\begin{table}
    \caption{Log likelihood ratios of estimated power law distributions compared with other candidate distributions.} 
    \begin{center}
    \begin{tabular}{| p{1.0cm}|| p{1.5cm} | p{1.5cm} || p{1.5cm} | p{1.5cm} |} \hline
    \textbf{}      	& \multicolumn{2}{c||}{exponential} & \multicolumn{2}{c|}{stretched exp} \\ \hline
    \textbf{} 		& \(\mathcal{R}\) & \(p\) & \(\mathcal{R}\) & \(p\) \\ \hline
    \textbf{PG} 	& \(2.7 \times 10^3\)  & \(\mathcal{O}(10^{-100})\) & 34 & \(9 \times 10^{-4}\) \\ \hline 
    \textbf{APG} 	& \(7.1 \times 10^4\)  & 0 	    & \(1.0 \times 10^3\) & \(\mathcal{O}(10^{-85})\) \\ \hline

    \end{tabular}
    \end{center}

    \begin{center}
    \begin{tabular}{| p{1.0cm}|| p{1.5cm} | p{1.5cm} || p{1.5cm} | p{1.5cm} |} \hline
    \textbf{}      	& \multicolumn{2}{c||}{lognormal} & \multicolumn{2}{c|}{powerlaw with cutoff} \\ \hline
    \textbf{} 		& \(\mathcal{R}\) & \(p\) & \(\mathcal{R}\) & \(p\) \\ \hline
    \textbf{PG} 	& -3.4  & 0.10 & -8.5 & \(3.9 \times 10^{-5}\) \\ \hline 
    \textbf{APG} 	& -2.3  & 0.26 & -5.3 & \(1.1 \times 10^{-3}\) \\ \hline

    \end{tabular}
    \end{center}    
    
    \label{tbl:m_loglike}
    
    \bigskip
    
\small
The log likelihood ratio, \(\mathcal{R}\), and their associated \(p\)-values, \(p\), for fits of four alternative distributions compared with the  fit of the power law distribution. Statistics were gathered for PG, planar growth with \(m=2\), \(N=10^4\), and APG, Apollonian planar growth with \(N=5 \times 10^5\). Positive values of \(\mathcal{R}\) indicate that the powerlaw hypothesis is the preferred model of the data, \(p\) is the significance value of the log likelihood ratio.

\end{table}

\section{Analysis of Planar Growth}
\label{sec:analysis}

Having investigated the degree distribution we now take a look at other key indicators of global structure. We begin with the small world property and assortativity. Subsequently we examine how the planarity constraint affects the distribution of angles between edges.

\subsection{The small world property and assortativity}
\label{sec:small_world}

\begin{figure}
  \begin{center}
    \includegraphics[width=0.95\textwidth]{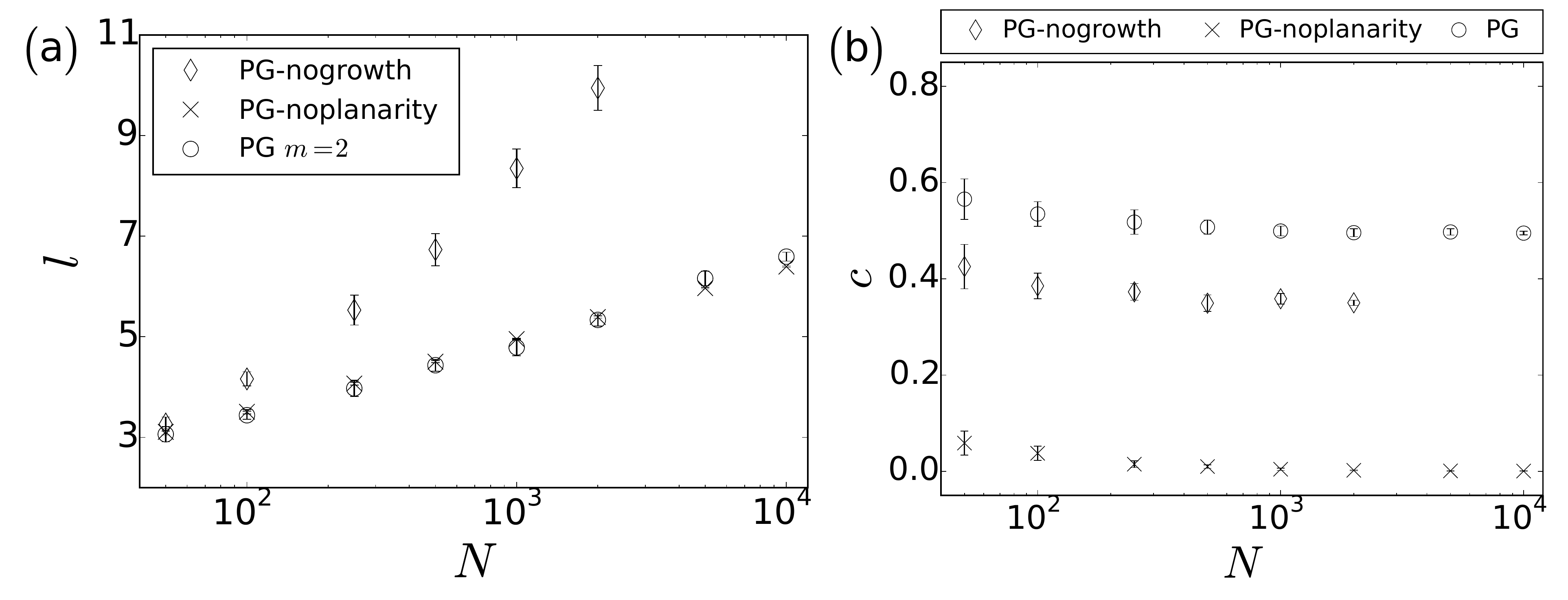}
  \end{center}
  \caption{(a) Mean characteristic path length for PG, PG-nogrowth and PG-noplanarity networks with varying order \(N\). Note the logarithmic scaling on the x-axis. (b) Clustering for the same networks. Error bars in each image is one standard deviation.}
  \label{fig:sw_test}
\end{figure}

We seek to determine if PG networks have the small world property; the defining characteristics of which are that the network's clustering coefficient \cite{watts1998collective}, \(c\), is high and the network's mean characteristic path length, \(l\), scales with \(N\) as \(l \sim \text{ln} \: N\). Here, \(l = \sum_{i, j \in V} d(i, j) / N(N-1) \) with \(V\) the set of vertices of the network and \(d(i, j)\) the length of the shortest topological path between \(i\) and \(j\). For random scale-free networks with \(2 < \alpha < 3\) it is known that \(l\) scales with \(N\) as follows: \(l \sim \text{ln} \: \text{ln} \, N\) \cite{cohen2003scale}. However the order of the networks, \(N = 10^4\), does not permit the precision necessary to confirm if this is the case for PG. Instead, figure \ref{fig:sw_test}a, a plot of the observed \(l\) for PG networks with varying \(N\) and \(m = 2\), indicates that \(l\) grows roughly logarithmically with \(N\).

Figure \ref{fig:sw_test}b shows that, for the same network, clustering is high for \(m = 2\). Large values of \(c\) in this case are accounted for by the fact that when a node is added it will form connections with the end nodes of nearby edges. PG networks with \(m \ge 2\) will therefore tend to form triangles with nearby edges. Furthermore, nearby edges deny a significant portion of the network to new nodes, thereby exacerbating this tendency. On this basis we hypothesise that the planarity constraint induces high clustering in PG networks where \(m \ge 2\). Further results, not presented here, confirm that this hypothesis is indeed correct and also indicate that the logarithmic scaling of the mean characteristic path length with \(N\) also holds when \(m \ge 2\). On this basis we conclude that this class of networks are small worlds.

Finally for this section we consider the assortativity coefficient, \(a\), which we define, following Newman \cite{newman2002assortative}, as a correlation coefficient of the degrees at either ends of an edge, i.e.,

\begin{align}
    a = \frac{N^{-1} \sum_i j_ik_i - [N^{-1} \sum_i \frac{1}{2} (j_i + k_i)]^2}
    {N^{-1} \sum_i \frac{1}{2} (j_i^2 + k_i^2) - [N^{-1} \sum_i \frac{1}{2} (j_i + k_i)]^2}
\end{align}

\noindent
where $j_i$, $k_i$ are the degrees of the vertices at the end of the $i^{\mathrm{th}}$ edge.

Specifically, we investigate how \(a\) varies with \(m\). For PG networks of order \(N = 10^4\), it decreases from \(a_{m=1} = -0.029 \pm 0.006\) to \(a_{m=3} = -0.066 \pm 0.002\), i.e., the networks are mildly disassortative and this tendency increases as \(m\) increases. Plots (not presented) of this relationship show it to be roughly linear. These results are in line with the well known fact that random scale-free networks are disassortative \cite{maslov2004detection, park2003origin}. A partial explanation of this phenomenon that has been offered is that there is a limited number of possible edges that can lie between high degree hubs \cite{maslov2004detection}. So, in general, a scale-free network must feature connections between high and low degree nodes. 

\subsection{Angle distribution}

\begin{figure}

  \begin{center}
    \includegraphics[width=0.95\textwidth]{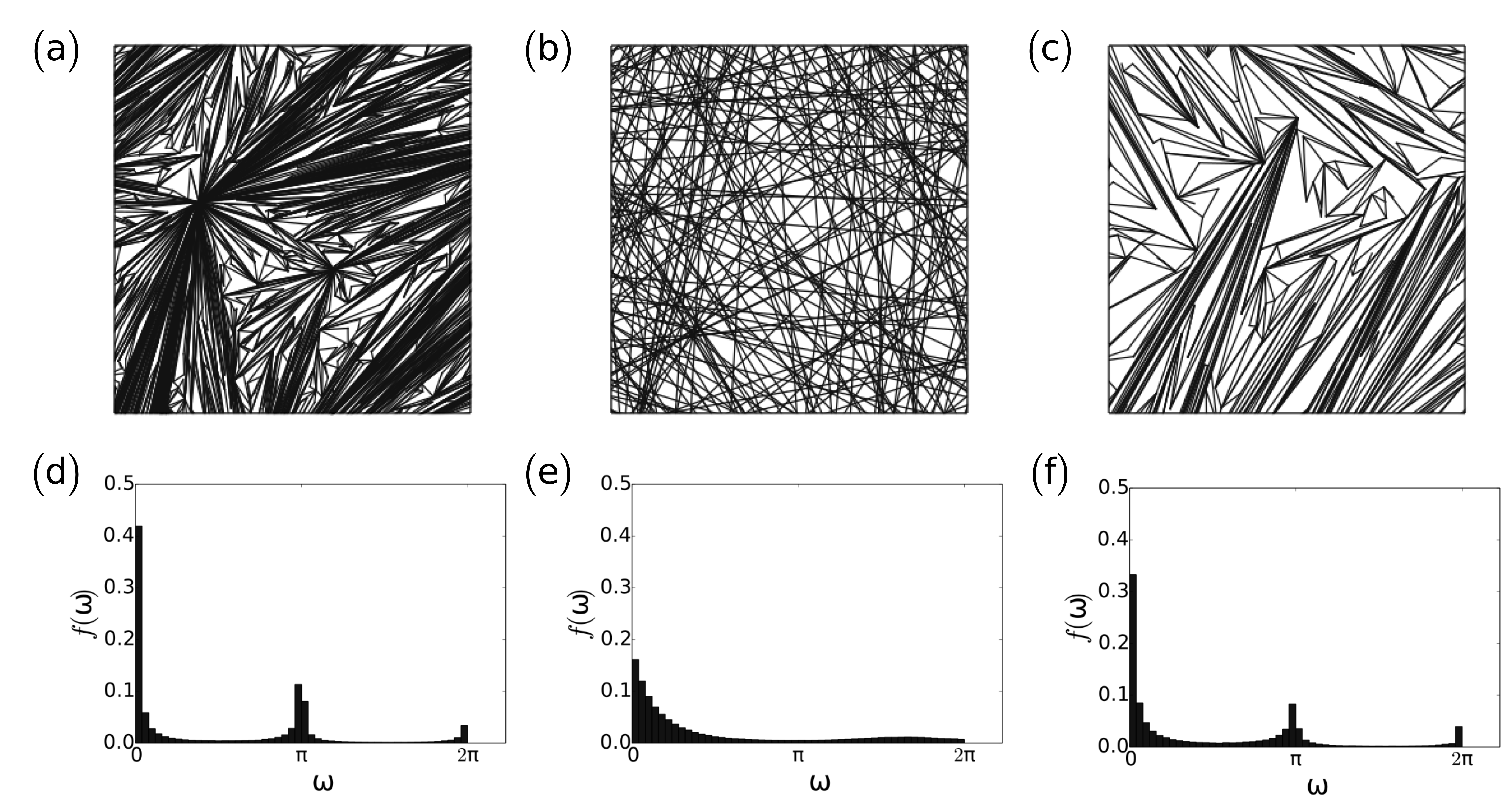}
  \end{center}

\caption{(a) Visualisation of a \(0.35 \times 0.35\) patch of a \(N = 10^4, m = 2\) planar growth network. (b) Visualisation of a \(0.01 \times 0.01\) patch of a \(N = 10^4, m = 2\) \(\text{PG-noplanarity}\) network. (c) Visualisation of a \(0.35 \times 0.35\) patch of a \(N = 2 \times 10^3, m = 2\) \(\text{PG-nogrowth}\) network. (d), (e) and (f) the probability mass for the angle, \(\omega\), between successive clockwise edges at a node of the network immediately above.}
\label{fig:vis_angle}
\end{figure}

Visualisations of PG, \(\text{PG-nogrowth}\) and \(\text{PG-noplanarity}\) networks are shown in figure \ref{fig:vis_angle}. A notable qualitative feature of the PG and \(\text{PG-nogrowth}\) plots are that edges emanating from the same node often closely bunch together. Combined with the observation of high clustering, this suggests that the space is predominately characterised by triangles with at least one highly acute angle. By contrast the \(\text{PG-noplanarity}\) network looks markedly different to the naked eye with edges crossing each other freely. 

Within city science, the distribution of angles between edges has been successfully employed to gain quantitative insight into road networks \cite{chan2011urban, barthelemy2013self}. In a similar fashion we here consider those edges incident to a vertex in clockwise order and calculate the angle \(\omega\) between subsequent pairs. The probability density of \(\omega\) is presented as a series of histograms beneath the corresponding visualisations in figure \ref{fig:vis_angle}. In the planar growth case three peaks are apparent; at zero, \(\pi\) and \(2\pi\) radians. The peak at zero is the largest and indicates the high number of acute angles just described. The peak at \(2\pi\) is evidence that in some cases the acute angle will be complemented by a large angle. The difference between the peaks at zero and \(2\pi\) indicates that in many cases several acute angles will be recorded at a single node in a fan like structure. There will also be occasions when these fan like structures are formed next to pre-existing edges. When this happens the fan will spread towards the edge without crossing it, thereby resulting in two edges incident at the same node that form an almost straight line. It is this phenomenon that accounts for the peak at \(\pi\). It should also be noted that the peaks at \(\pi\) and \(2\pi\) will be influenced by the boundary conditions; fans that form near the corners will contribute to the peak around \(2\pi\) while those that appear next to the middle of a side will contribute to the peak at \(\pi\).

The \(\text{PG-noplanarity}\) histogram shows a large proportion of small angles in a distribution that smoothly and rapidly tails off. There is a small bump at higher values of \(\omega\) which is a consequence of the square's boundary, i.e., nodes at the corners will tend to have some \(\omega > 3\pi / 4\). Finally, the \(\text{PG-nogrowth}\) histogram is very similar to that of the PG networks, confirming that the angular structure is a consequence of the planarity constraint.

\begin{table}
    
    \caption{Estimated exponents of crossing probability networks.} 
    \begin{center}
    \begin{tabular}{| p{0.2cm} | p{0.6cm} | p{0.6cm}| p{0.6cm}| p{0.6cm} | p{0.6cm}| p{0.6cm}| p{0.6cm} | p{0.6cm}| p{0.6cm}| p{0.6cm} | p{0.6cm}|} \hline
    \textbf{\(\chi\)} & 0.0 & 0.1 & 0.2 & 0.3 & 0.4 & 0.5 & 0.6 & 0.7 & 0.8 & 0.9 & 1.0 \\ \hline
    \textbf{\(\alpha\)} & 2.83 & 2.89 & 3.00 & 3.13 & 3.31 & 3.55 & 3.95 & 4.68 & 6.21 & 5.71 & 6.25\\ \hline
    \textbf{\(\sigma\)} & 0.01 & 0.01 & 0.02 & 0.01 & 0.02 & 0.04 & 0.03 & 0.08 & 0.32 & 0.07 & 0.05 \\ \hline

    \end{tabular}
    \end{center}
    \label{tbl:CP_alpha}
    \bigskip
\small
Each exponent \(\alpha\) is estimated assuming a power law degree distribution for a batch of twenty experiments which grow an \(N = 10^4\), \(m = 2\) network with the crossing probability \(\chi\) that is specified in the first row. Standard error, \(\sigma\), is reported in the third row.
\end{table}

\newpage

\section{Planarity relaxation}
\label{sec:planar:relax_planarity}

\begin{figure}
  \begin{center}
    \includegraphics[width=0.44\textwidth]{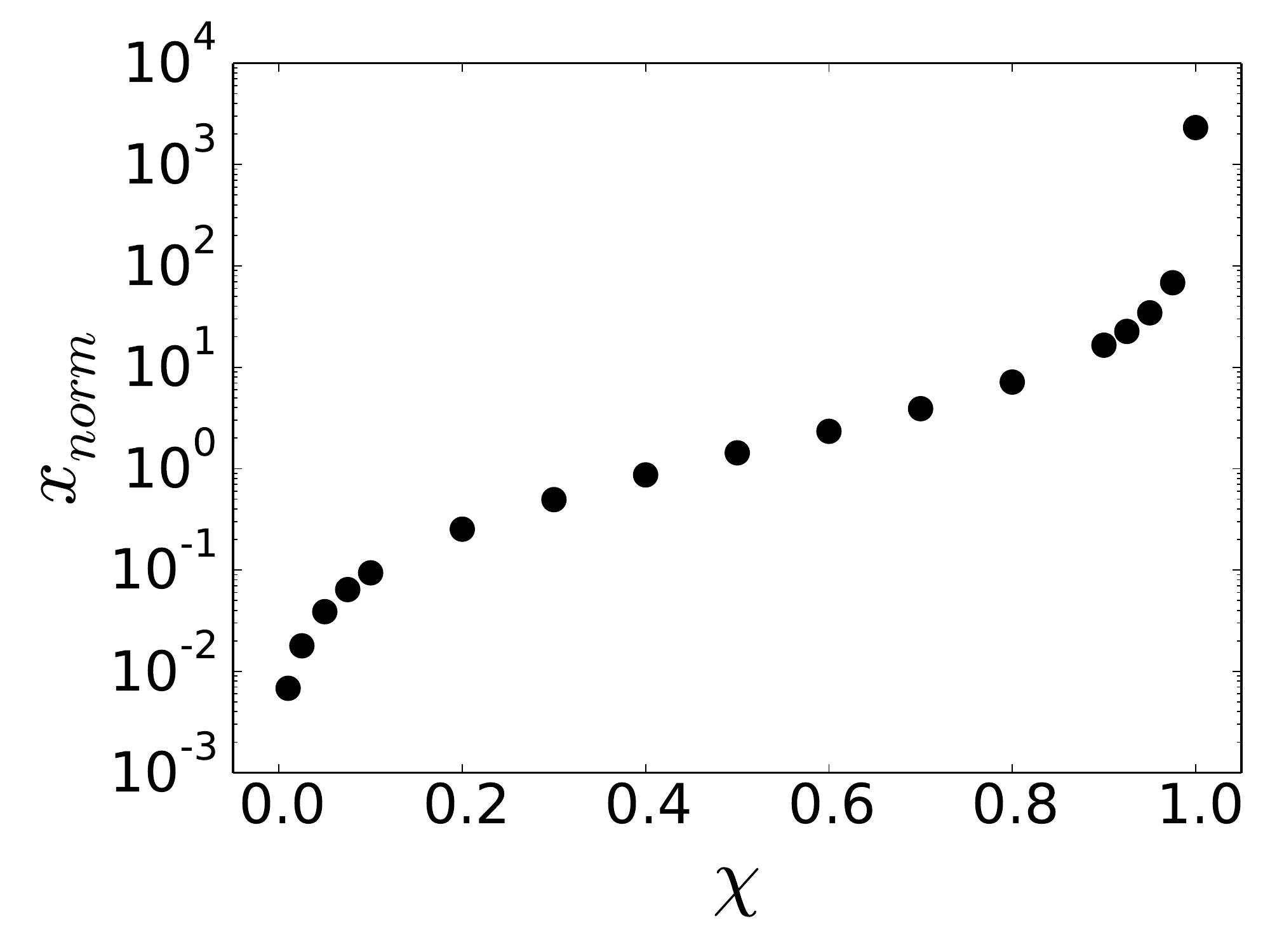}
  \end{center}
  \caption{The normalised count of crossings, \(x_{norm}\), observed in experiments with varying \(\chi\). Normalisation was observed by dividing \(x\), the number of crossings, by 20,020, the number of edges. The true value of \(x_{norm}\) when \(\chi = 0\) cannot be represented on logarithmic axes and has been approximated by the value for \(\chi = 0.01\).}
  \label{fig:CP_norm}
\end{figure}

The contrast between the PG and the \(\text{PG-noplanarity}\) degree distributions is dramatic and we would like to investigate intermediate networks. To do so we introduce a new parameter; \(\chi \in [0, 1] \), the crossing probability. This parameter is applied in step 2b of the PG algorithm where, instead of rejecting crossings outright, we allow them with probability \(\chi\). We grow networks with \(N = 10^4\) and \(m = 2\) while using a different value of \(\chi\) in the range 0.0 and 1.0 for each experiment. Our first result, presented in figure \ref{fig:CP_norm}, is a plot of the normalised number of crossings which shows that the number of crossings increases in a roughly exponential fashion between \(\chi = 0.1\) and \(\chi = 0.9\). Beyond \(\chi = 0.9\) the number of crossings increases significantly in comparison to the previous regime.

\begin{figure}
  \begin{center}
    \includegraphics[width=0.95\textwidth]{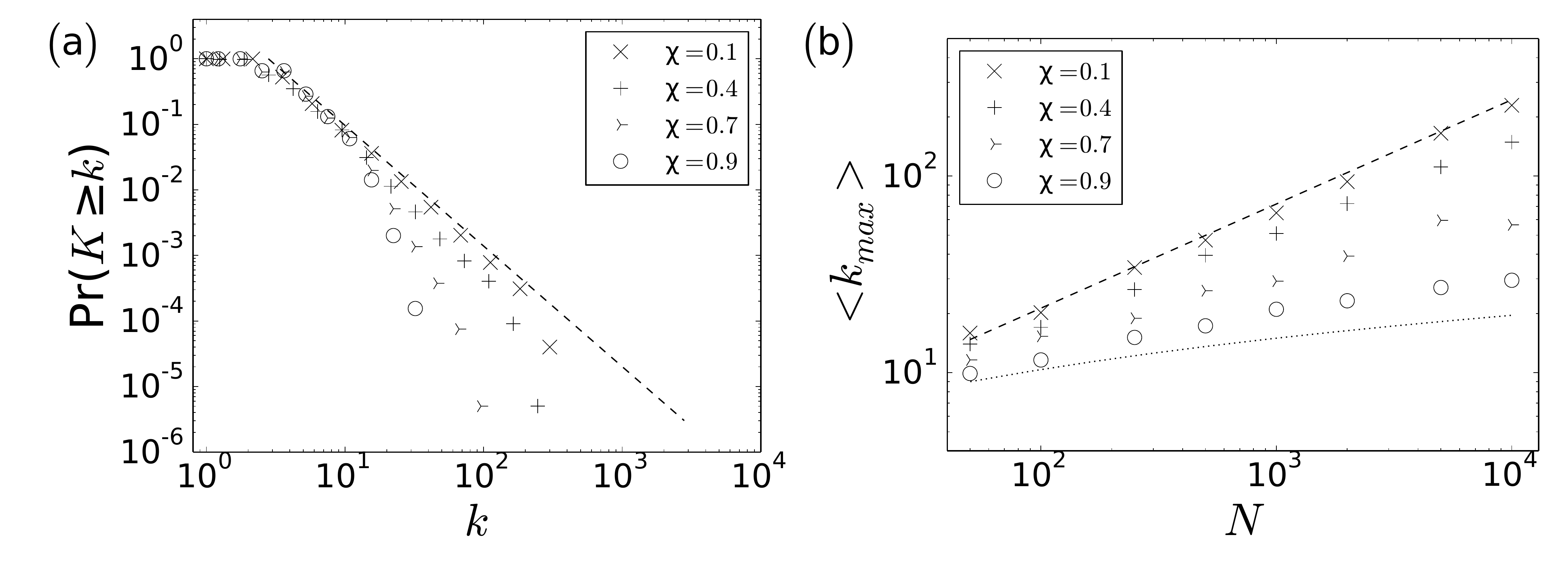}
  \end{center}

  \caption{(a) Cumulative degree distributions for networks created using planar growth with a probability \(\chi\) of accepting edge crossings. The dashed line is the power law with exponent \(\alpha_{m = 2}\), the best fit for the \(\chi = 0.0\) experiment. (b) Average maximum degree observed in the same experiments. Dotted and dashed lines are the same references plotted in figure \ref{fig:m2_mechanisms_k_hist} and are fits for the \(\chi = 0.0\) cases and \(\chi = 1.0\) cases respectively.}
  \label{fig:cp_degree_dist}
  
\end{figure}

The associated degree distributions are shown in figure \ref{fig:cp_degree_dist}a where we see a smooth transition from a power law to an exponential curve as \(\chi\) increases from 0.0 to 1.0. Similarly, in figure \ref{fig:cp_degree_dist}b we present the average maximum degree where plots for low \(\chi\) match the predicted maximum of a power law while increasing \(\chi\) leads to curves that more closely match the exponential prediction. Taken together, this evidence shows a smooth transition from a heavy tailed to an exponential degree distribution as \(\chi\) increases. We also estimated exponents, assuming a power law distribution, and report the results in table \ref{tbl:CP_alpha}, finding an increasing trend for the exponent with \(\chi\) for networks with \(\chi \le 0.8\). However, from an examination of figure \ref{fig:cp_degree_dist}a, it is clear that networks for which \(\chi > 0.7\) have a degree distribution that is exponential and we can therefore disregard exponents in this region of the parameter space.

\begin{figure}
\centering
  \begin{center}
    \includegraphics[width=0.95\textwidth]{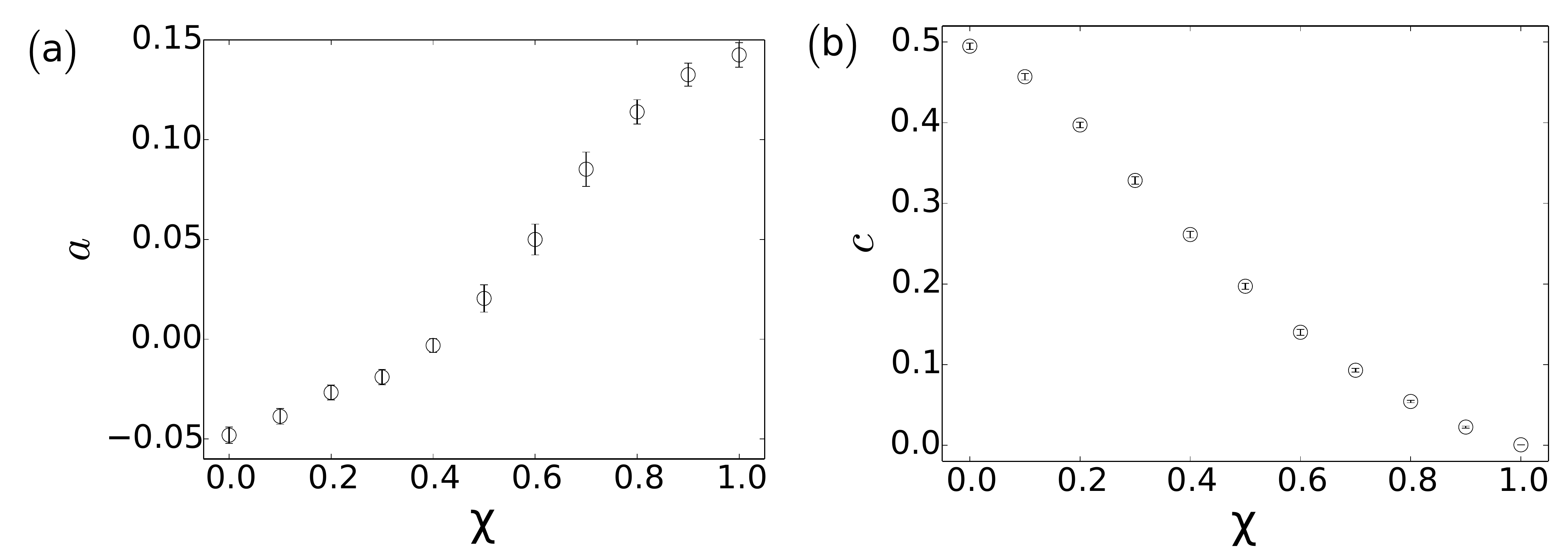}
  \end{center}

\caption{(a) Average assortativity observed in PG networks with varying \(\chi\). (b) Average clustering observed in the same experiments. Each data point relates to twenty networks grown using \(n = 10^4\), \(m = 2\). Error bars represent one standard deviation.}

\label{fig:cp_measures}
\end{figure}

Assortativity is plotted in figure \ref{fig:cp_measures}a and again exhibits a smooth transition, this time from mild disassortativity to assortativity. As has been discussed in section \ref{sec:small_world}, preventing edge crossing results in nodes being more likely to connect to nodes at the ends of nearby edges. We can calculate, for node \(i\), the strength, \(s_i = \sum_{j \in V(i)} w_{ij}\) where \(V(i)\) is the set of vertices connected to \(i\) and \(w_{ij}\) is the Euclidean distance between \(i\) and \(j\). Nodes with high strength, i.e., those whose edges have a high total length, will be favoured. Such nodes will tend to attract connections from new, low degree nodes thus accounting for the disassortativity. Clearly this tendency will be relaxed as \(\chi\) increases leading to more assortative networks.

We consider the clustering of these networks in figure \ref{fig:cp_measures}b noting a high \(c_{\chi = 0.0} = 0.49\) descending to a negligible value for \(c_{\chi = 1.0}\). High clustering occurs for \(\chi = 0.0\) for the reasons outlined in section \ref{sec:small_world}. On the other hand a new node connects freely to any existing node in the \(\chi = 1.0\) case and, hence, this model displays no clustering, equivalent to the uniform attachment model.

\section{Comparison of APG with existing Apollonian growth}
\label{sec:apg:comparison}

In section \ref{sec:apg} we introduced APG as a refinement to the Random Apollonian Network noting that the exponent of its degree distribution was \(\alpha_{\text{APG}} = 2.77 \pm 0.01\). We contrast this with the analytically derived exponents for the original Apollonian network, \(\alpha_{DAN} = 2.585\), and the RAN, \(\alpha_{RAN} = 3.0\). The APG's exponent lies between these two values and we contend that this is because APG can be thought of as a generalisation of the two existing models.

\begin{figure}
\centering
    \includegraphics[width=.9\linewidth]{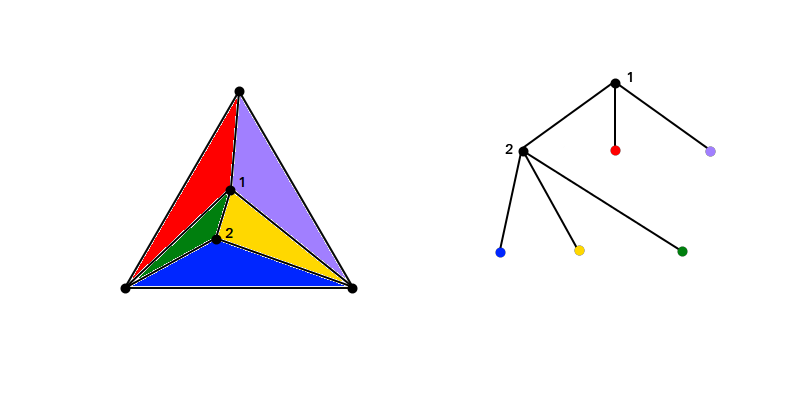}
\caption{A triangulation resulting from an Apollonian growth process represented as a ternary tree. Here a root node, 1, and one subsequent node, 2, have been added to the triangulation, resulting in five faces. This triangulation is represented by a ternary tree on the right hand side. The internal nodes of the tree correspond to the nodes of the triangulation and have been numbered as such. The leaves of the tree correspond to the faces and the colouring scheme indicates this.}
\label{fig:triangulation}
\end{figure}

A triangulation created by any of the three Apollonian growth processes can be represented as a ternary tree where the internal nodes of the tree correspond to nodes of the network and leaves of the tree to the triangular faces \cite{albenque2008some}, see figure \ref{fig:triangulation}. In the case of the RAN, picking faces of the triangulation uniformly at random is equivalent to picking leaves of the tree uniformly at random. Therefore, the corresponding tree for the RAN will tend to grow in depth since leaves at the bottom of the tree will appear in greater abundance.

DAN constructions begins with \(K_4\) embedded in \(\mathbb{R}^2\). Growth is an iterative process where, at each stage, a new node is placed within each of the graph's internal faces. Each of these new nodes is then connected to the vertices of its containing face resulting in three new faces. This recursive splitting of the triangle is repeated \(t\) times and the corresponding ternary tree has depth \(t + 2\) and is both full and complete. Most importantly, the tree for the DAN is shallower than that of the RAN.

The consequence for the degree distributions of the triangulations is as follows: at depth \(t\) there are \(3^t\) potential nodes of which \(3 \times 2^{t-1}\) will connect to the triangulation's root node. The triangulation associated with the DAN is guaranteed to fill those locations that maximise the degree of the root node. Furthermore, note that the structure of the ternary tree is self-similar. As such any node within the DAN receives the maximum number of connections from its descendants on the tree. Meanwhile, degree in the RAN will be distributed more evenly since the new additions at greater depths will not have the same tendency to link to those at shallower levels of the tree. We therefore expect the DAN to exhibit a heavier tail in its degree distribution and this accounts for the fact that \(\alpha_{DAN} < \alpha_{RAN}\).

In the case of APG there will be a tendency to place nodes within those faces with the greatest area. The intuition here is that the earlier a face is created the larger its area will be and therefore those faces that are at a shallow depth within the ternary tree will be favoured for selection. On the other hand, as more nodes are added, a greater proportion of the triangle's total area is covered by newer triangles at greater depth and these will come to be favoured over time. Therefore, nodes in the APG's ternary tree appear at a depth between those of the DAN and the RAN and this in turn explains why we see an intermediate exponent for the degree distribution.

\subsection{Area weighting and trisection}
\label{sec:apg:weight_exp}

We test the hypothesis of the previous section by varying the extent to which face area influences its selection within APG. To this end the equation \ref{eqn:apg} is modified as follows:

\begin{align}
\pi_i(t) = \frac{a_i^{\beta}}{\sum\limits_{j \in F_t} a_j^{\beta}},
\label{eqn:apg_weighted}
\end{align}

\noindent where \(\beta\), the area weighting exponent, is a parameter controlling the influence of a given triangle's area. Clearly, as \(\beta \rightarrow 0\), faces will be chosen at random and RAN will be recovered. Conversely, when \(\beta \rightarrow \infty\) larger faces will be favoured and we expect, from the arguments preceding, to recover the DAN instead.

In table \ref{tab:beta_vs_alpha} we report the exponents for networks created using this variation of the APG and varying \(\beta\). For values of \(\beta < 1\) we see precisely the result predicted, as \(\beta \rightarrow 0\), \(\alpha \rightarrow 3\). On the other hand, for values of \(\beta > 1\), a saturation effect has taken hold and the exponent remains around 2.76. This contradiction with the predicted behaviour occurs because we have assumed that shallow faces in the ternary tree will always have a greater area than deeper ones. Since nodes are placed randomly upon their containing triangle, this isn't necessarily the case in Apollonian Planar Growth. Thus, nodes tend to appear deeper in the ternary tree than our initial hypothesis assumed.

\begin{minipage}{\linewidth}
\centering
\bigskip
\captionof{table}{Variation of the degree distribution with area weighting.} \label{tab:beta_vs_alpha}
    \begin{tabular}{| p{0.25cm} | p{0.5cm} | p{0.5cm} | p{0.5cm} | p{0.5cm} | p{0.5cm} | p{0.5cm} | p{0.5cm} |}
    \hline
    \textbf{\(\beta\)} & \(10^{-3}\) & \(10^{-2}\) & \(10^{-1}\) & \(10^{0}\) & \(10^{1}\) & \(10^{2}\) & \(\infty\) \\ \hline
    \textbf{\(\alpha\)} & 2.92 & 2.93 & 2.82 & 2.76  & 2.77 & 2.76 & 2.77 \\ \hline
    \end{tabular}\par
\bigskip
Estimated values of the exponent of the degree distribution, \(\alpha\), observed for APG networks of order \(10^5\) with varying area weighting exponent, \(\beta\). The standard error, \(\sigma\), in each case is 0.01.
\bigskip
\end{minipage}

\begin{figure}
  \begin{center}
    \includegraphics[width=0.58\textwidth]{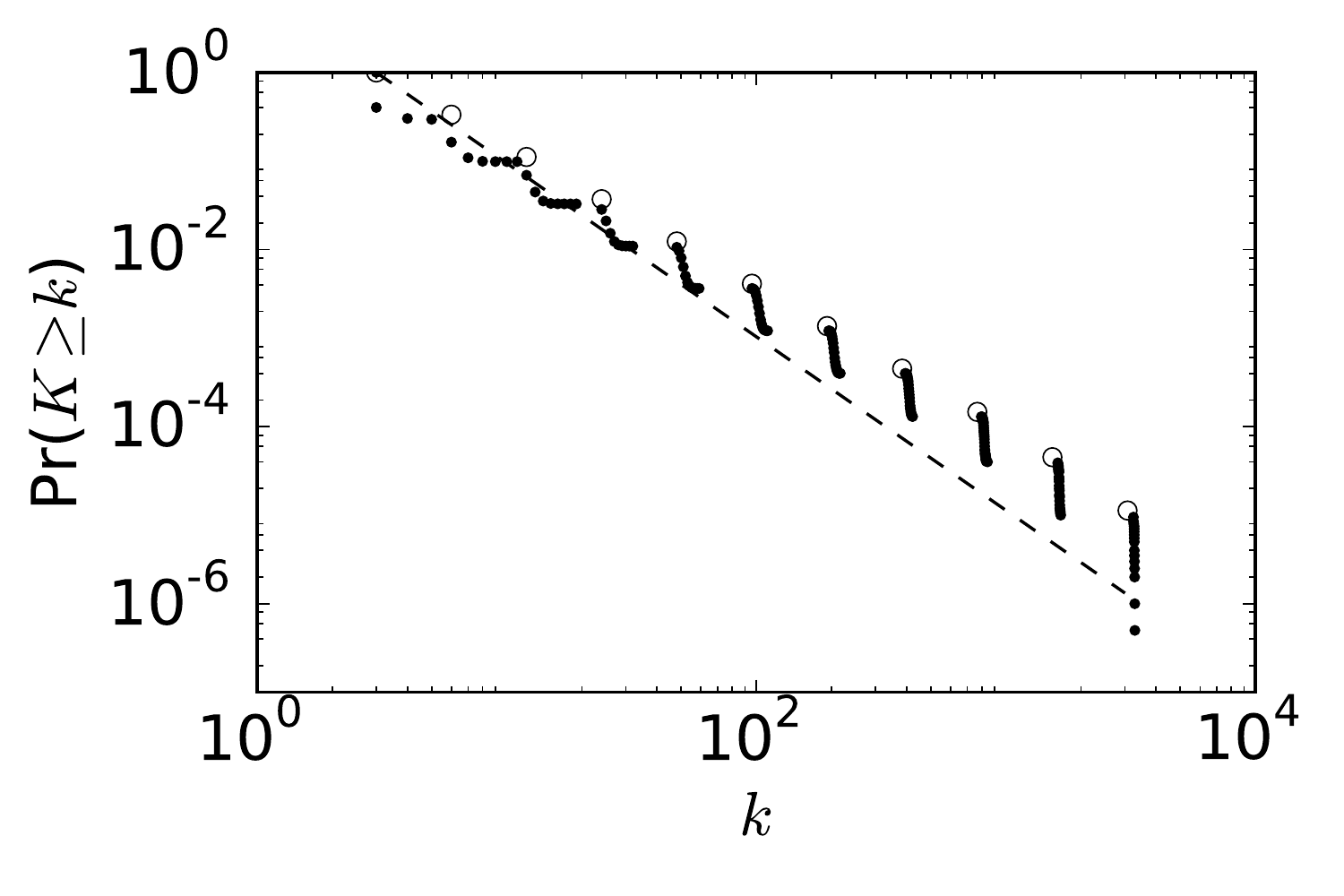}
  \end{center}
  \caption{Black dots are the degree distribution of a network of order \(n = 10^5\) grown using Apollonian Planar Growth with \(\beta = 100\). Faces were divided by trisecting in to three equal areas in this version of the model. Empty circles are the degree distribution of an Apollonian network of the same order. Dashed line is a plot the best fit of the exponent, \(\alpha_{\beta = 100} = 2.85 \pm 0.01.\)}
  \label{fig:trisection_100}
\end{figure}

In light of this reasoning we further modify the algorithm by placing each new node so that it exactly trisects its containing face; thereby guaranteeing a hierarchy of face sizes by depth within the tree. For \(\beta < 1\) behaviour was again as expected; exponents were observed to increase from \(2.93 \pm 0.01\) to \(2.68 \pm 0.01\) as \(\beta\) increased from \(10^{-2}\) to \(10^{-0.5}\). A further experiment with \(\beta = 10^2\) gave an exponent of \(2.85 \pm 0.01\) which, prima facie, suggests that the hypothesis is incorrect. However, it is apparent from figure \ref{fig:trisection_100} that the fit is not indicative of the degree distribution of the area weighted APG with trisection. This is because the Clauset et al.\ method is inappropriate for quantifying the exponent of power laws that exhibit the sort of discretisation we see in the plot.

To better understand the distribution as \(\beta \rightarrow \infty\) we instead follow the formula for the degree distribution of a DAN presented in Andrade et al's original paper \cite{andrade2005apollonian} and plot it on figure \ref{fig:trisection_100} alongside our own data. It is clear that the discretisation of the experiment closely matches that of the analytical calculation. In a further experiment we set \(\beta = \infty\), i.e. the largest triangle was always chosen, and grew a network of 265,720 nodes, the order of an Apollonian network that has been iterated 11 times. In this case the analytical calculation exactly matches the experimental data, confirming that the area weighted APG with trisection approximates the DAN as \(\beta\) increases.

To complete the analysis we considered networks with \(\beta < 0\), results obtained indicate that an exponential distribution takes hold in this regime. In this regime, new nodes tend to appear within the model's smallest triangle, thereby creating new smallest face from the resulting trisection. Thus, new nodes will tend to congregate in the same region of the model. This contrasts with the \(\beta \ge 0\) case where division of the largest face effectively lessens the probability of that region being selected in the next iteration of the process. Thus, the potential for nodes to be distributed over the entire face is a key feature in the onset of the power law degree distribution.

\section{Summary}
\label{sec:summary}

We have introduced planar growth as a model of spatial network formation in which a network is grown over time such that planarity is maintained. Resulting networks have been found to be scale-free, have the small world property and are mildly disassortative. It should be noted that PG attains the power law degree distribution with a uniform distribution of nodes in space. As far as we are aware this is only the second example, Mukherjee \& Manna \cite{mukherjee2006weighted} being the first, of a spatial growth process that attains this outcome under this constraint. The scale-free property is dependent on two aspects of the process; sequential growth and maintenance of planarity. Removal of either aspect results in a network with an exponential degree distribution.

To the extent that the planarity constraint is relaxed the degree distribution degrades from a power law to the exponential case. Similarly, smooth crossover was noted for the clustering and assortativity of these networks. We have discussed those spatial networks that are nearly planar and further note that Newman has articulated a desire for a quantification of the degree of planarity \citep{newman2010networks}. We offer these results as an intial step towards resolving this question.

A refinement of the model, Apollonian Planar Growth, demonstrated a connection between planar growth and Apollonian networks. Weighting the area selection of triangles during Apollonian Planar Growth allowed us to easily recover the Random Apollonian Network while a further variation, trisection, was required in order to produce the Deterministic Apollonian Network. As such, weighted Apollonian Planar Growth with trisection acts as a framework that generalises the two existing Apollonian models. 

This paper also opens up various interesting questions for future research.  In the research presented here we have analysed the effects of a planarity constraint on the structure of a growing network. It appears of interest to investigate planarity constraints in conjunction with other network formation mechanisms. For instance, an area in network research that has found much attention in the literature are questions of optimal design of network structures \cite{mathias2000small, i2003optimization, colizza2004network, gastner2006shape, brede2010coordinated}. It would be of interest to further investigate to what extent planarity restrictions can constrain such optimal network topologies.

\section{Acknowledgements}
This work was supported by an EPSRC Doctoral Training Centre grant (EP/G03690X/1). The authors acknowledge the use of the IRIDIS High Performance Computing Facility, and associated support services at the University of Southampton, in the completion of this work. The authors would also like to thank Massimo Stella for numerous helpful discussions at the inception of this project.


\bibliographystyle{unsrt}
%

\end{document}